\documentstyle[12pt,epsf]{article}
\input tcilatex
\QQQ{Language}{
British English
}

\begin{document}

\title{Some exact non-vacuum Bianchi VI$_0$ and VII$_0$ instantons}
\author{John D. Barrow$^1$\thanks{%
E-mail: J.D.Barrow@sussex.ac.uk} , Yves Gaspar$^1$\thanks{%
E-mail: yfj-mg@star.cpes.susx.ac.uk} \& P. M. Saffin$^2$\thanks{%
E-mail: P.M.Saffin@damtp.cam.ac.uk} \\
%EndAName
$^1$Astronomy Centre,University of Sussex, \\
Brighton BN1 9QJ, UK\\
$^2$DAMTP, Silver Street,\\
Cambridge CB3 9EW, UK}
\date{\today }
\maketitle

\begin{abstract}
We report some new exact instantons in general relativity. These solutions
are K\"ahler and fall into the symmetry classes of Bianchi types VI$_0$ and
VII$_0$, with matter content of a stiff fluid. The qualitative behaviour of
the solutions is presented, and we compare it to the known results of the
corresponding self-dual Bianchi solutions. We also give axisymmetric Bianchi
VII$_0$ solutions with an electromagnetic field.
\end{abstract}

\renewcommand{\topfraction}{0.8} 

%@twocolumnfalse\endcsname

%\pacs{PACS numbers: 98.80.Cq }

%\vskip2pc]

%%%%%%%%%%%%%%%%%%%%%%%%%%%%%%%%%%%%%%%%%%%%%%%%%%%%%%%%%%%%%%%%%%%%%%%

\section{ Introduction}

Solutions to field equations in imaginary time have been of interest for
many years and early work formalised the relevance of such systems to
tunnelling problems \cite{mclaughlin72} \cite{coleman77}. Much of the
current interest in instantons dates from the work of Belavin, Polyakov,
Shvarts, and Tyupkin (BPST) \cite{bpst75}, where an exact solution to the
SU(2) Yang Mills equations in imaginary time was found. The relevance of
these solutions was then clarified by t'Hooft \cite{thooft76} and by Jackiw
and Rebbi \cite{jackiw76} who showed that the BPST instanton solution is a
mediator for the transition between different vacua of the gauge theory.

The BPST instanton was found by looking for solutions with (anti) self-dual
field strength. This led to the suggestion that analogous solutions may
exist in gravity \cite{belinskii78}, where the duality governs the curvature
two-form. The usefulness of such a duality is that it not only makes the
curvature automatically solve Einstein's vacuum equations \cite{eguchi80},
but also reduces the equations to first order. A geometry which possesses a
symmetry also simplifies the problem. An important class of symmetries is
the Bianchi classification of spatially homogeneous 3-geometries\cite{ryan}.
Some solutions of this type with the property of self duality are known and
can be found in reference \cite{lorenzpetzold89}; we shall extend the
collection of known solutions by finding metrics with the symmetry of
Bianchi types VI$_0$ and VII$_0$ that satisfy Einstein's equations with a
stiff fluid, that is where the pressure equals the energy density. This
means that we cannot use the simplification of self duality, however we do
use it as a guide to find an initial ansatz. We shall also see that this
ansatz allows a natural complex structure such that the metric is  K\"ahler.

The importance of gravitational instantons emerges in the path-integral
formalism of quantum gravity. In this approach one  considers a weighted sum
over all Euclidean metrics, where the weight is the exponential of the
Euclidean action. Instantons, as solutions to the classical field equations,
are expected to provide the dominant contribution to such a sum, and this
allows a saddle-point approximation to be used in the neighbourhood of these
solutions.

The organisation of the paper is as follows. First we define the system of
equations to be solved, and describe the self-dual vacuum solution. We solve
the Bianchi VII$_0$ Einstein equations in the presence of a stiff fluid.
Next, we follow Misner \cite{misner} by writing the action in the familiar
form of a particle in a potential, in order to describe the qualitative
nature of our solution. The Bianchi VI$_0$ solution follows from the Bianchi
VII$_0$ results.

%%%%%%%%%%%%%%%%%%%%%%%%%%%%%%%%%%%%%%%%%%%%%%%%%%%%%%%%%%%%%%%%%%%%%%

\section{The Bianchi VII$_0$ equations}

The Bianchi universes are four-manifolds with embedded three geometries
which are spatially homogeneous but anisotropic; they each possess a
three-dimensional, simply-transitive, isometry group \cite{ryan}. To exploit
this isometry we write the metric using the left-invariant one-forms of the
symmetry group, $\sigma ^a$; as we are interested in instantons, we use a
Riemannian signature, 
\begin{eqnarray}
{\rm ds}^2={\rm d}\bar \tau ^2+a^2(\bar \tau )(\sigma ^1)^2+b^2(\bar \tau
)(\sigma ^2)^2+c^2(\bar \tau )(\sigma ^3)^2.
\end{eqnarray}
The left-invariant one-forms satisfy the structure equations, 
\begin{eqnarray}
{\rm d}\sigma ^a=\frac 12C_{\;bc}^a\sigma ^b\wedge \sigma ^c,
\end{eqnarray}
For the case of Bianchi VII$_0$, the structure constants take the values 
\begin{eqnarray}
C_{\;32}^1=-C_{\;23}^1=C_{\;13}^2=-C_{\;31}^2=1.  \label{b7struct}
\end{eqnarray}
These metrics  support Killing vectors (right-invariant vector fields) with
the same structure constants as the Euclidean group in two dimensions. The
self-dual solution will be of use to us in constructing an ansatz, so we
shall calculate the torsion-free connection forms, $\Theta _{\;b}^a$ \cite
{eguchi80}, assuming they are metric connections, \mbox{$\Theta_{ab}=-%
\Theta_{ba}$}.Thus, we rewrite the metric in terms of a new time variable %
\mbox{$\rm{d}\bar{\tau}=abc\rm{d}\tau$} and introduce an orthonormal basis, 
\begin{eqnarray}
{\rm ds}^2 &=&a^2(\tau )b^2(\tau )c^2(\tau ){d}\tau ^2+a^2(\tau
)(\sigma ^1)^2+b^2(\tau )(\sigma ^2)^2+c^2(\tau )(\sigma ^3)^2  \label{orth}
\\
\omega ^0 &=&a(\tau )b(\tau )c(\tau ){\rm {d}\tau ,\;\;\;\;\omega ^1=a(\tau
)\sigma ^1,\;\;\;\;\omega ^2=b(\tau )\sigma ^2,\;\;\;\;\omega ^3=c(\tau
)\sigma ^3.}  \nonumber
\end{eqnarray}
We find the following non-vanishing connection forms: 
\begin{eqnarray}
\theta _{\;1}^0 &=&-\alpha ^{\prime }/(abc)\omega ^1,\;\;\;\theta
_{\;2}^1=-(a^2+b^2)/(2abc)\omega ^3  \nonumber  \label{connections} \\
\theta _{\;2}^0 &=&-\beta ^{\prime }/(abc)\omega ^2,\;\;\;\theta
_{\;3}^1=(a^2-b^2)/(2abc)\omega ^2 \\
\theta _{\;3}^0 &=&-\gamma ^{\prime }/(abc)\omega ^3,\;\;\;\theta
_{\;3}^2=-(a^2-b^2)/(2abc)\omega ^1,  \nonumber
\end{eqnarray}
where we have defined \mbox{$\alpha=\ln(a)$}, \mbox{$\beta=\ln(b)$}, %
\mbox{$\gamma=\ln(c)$} and $^{\prime }$ refers to differentiation with
respect to $\tau $. We now assume the connection to be self dual, which in
turn implies that the curvature is self dual \cite{eguchi80}, so %
\mbox{$\Theta_{01}=\Theta_{23}$} plus permutations. 
\begin{eqnarray}
\alpha ^{\prime }=-\frac 12(a^2-b^2),\;\;\;\beta ^{\prime }=\frac
12(a^2-b^2),\;\;\;\gamma ^{\prime }=\frac 12(a^2+b^2).
\end{eqnarray}
These equations show that $a(\tau )b(\tau )$ is a constant, and give a
metric, 
\begin{eqnarray}
{\rm {d}s^2=\frac{\lambda ^2}2\sinh (2\tau )\left( {d}\tau ^2+(\sigma
^3)^2\right) +\coth (\tau )(\sigma ^1)^2+\tanh (\tau )(\sigma ^2)^2,}
\label{sd_metric}
\end{eqnarray}
where $\lambda $ is an integration constant.

We now derive a non-vacuum solution, (which requires that self duality is
lost). However, we use the self-dual solution to lead us to the ansatz of %
\mbox{$a(\tau)=1/b(\tau)$}. In order to show that this solution does retain
the K\"ahler property, the metric is rewritten by introducing the complex
forms $\zeta $ defined by 
\begin{eqnarray}
\zeta ^1 &=&\omega ^0+i\omega ^3 \\
\zeta ^2 &=&\omega ^1+i\omega ^2,
\end{eqnarray}
where the $\omega ^\alpha $ are those defined in (\ref{orth}). The metric
now takes the form 
\begin{eqnarray}
{\rm {ds}^2=\zeta ^1\zeta ^{\overline{1}}+\zeta ^2\zeta ^{\overline{2}}.}
\end{eqnarray}
The K\"ahler form is then found to be 
\begin{eqnarray}
\Omega _K &=&ig_{a\overline{b}}\zeta ^a\wedge \zeta ^{\overline{b}} \\
&=&4abc^2{\rm {d}\tau \wedge \sigma ^3+4ab\sigma ^1\wedge \sigma ^2}
\end{eqnarray}
The closure of the K\"ahler form, \mbox{$\rm{d}\Omega_K=0$}, then requires %
\mbox{$(ab)'=0$}, so our ansatz leads to a K\"ahler manifold.

We now want to consider the full Bianchi VII$_0$ equations, without
requiring self duality. These can be found in refs. \cite{lukash73}, \cite
{lorenz84} where we are now using a Lorentzian signature, 
\begin{eqnarray}
{\rm ds}^2 &=&a^2(t)b^2(t)c^2(t){d}t^2\\
\nonumber &~&-a^2(t)(\sigma
^1)^2-b^2(t)(\sigma ^2)^2-c^2(t)(\sigma ^3)^2  \label{ein4} \\
\left( \ln (a^2)\right) ^{..}+a^4-b^4 &=&0  \label{aeqn} \\
\left( \ln (b^2)\right) ^{..}-a^4+b^4 &=&0  \label{beqn} \\
\left( \ln (c^2)\right) ^{..}-(a^2-b^2)^2 &=&0  \label{ceqn} \\
\left( \ln (abc)^2\right) ^{..} &=&4\left( (\ln c)^{.}(\ln ab)^{.}+(\ln
a)^{.}(\ln b)^{.}\right) +2\epsilon (abc)^2.  \label{int}
\end{eqnarray}
where $\epsilon \geq 0$ is the energy density, and $^{.}$ denotes
differentiation with respect to $t$. Now we take the hint from the self-dual
solution and assume \mbox{$a(t)=1/b(t)$}, but we do not take $\epsilon $ to
be zero. Then equations (\ref{aeqn},\ref{beqn},\ref{ceqn}) become, 
\begin{eqnarray}
\frac{{\rm d}^2 y}{{\rm d}x^2} &=&-\sinh (y)  \label{heqn} \\
\frac{{\rm d}^2 h}{{\rm d}x^2} &=&\sinh ^2(y/2),  \label{ay}
\end{eqnarray}
where we have used \mbox{$x=2t$}, \mbox{$a^2=\exp(y/2)$} and %
\mbox{$c^2=\exp(h)$}. Equation (\ref{heqn}) can be solved to give, 
\begin{eqnarray}
\cosh (y(x))=1+2\frac{k^2}{k^{\prime }{}^2}{\rm cn}^2(x/k^{\prime }+\alpha
\mid k^2),  \label{ysoln}
\end{eqnarray}
where $k$ is the modulus of the elliptic function, $k^{\prime }$ is the
complementary modulus, and $\alpha $ is an arbitrary constant. Our notation
conforms to that of \cite{abramowitz}. With (\ref{ysoln}) substituted into (%
\ref{heqn}) we find 
\begin{eqnarray}
\frac{{\rm d}^2h}{{\rm d}x^2}=\frac{k^2}{k^{\prime }{}^2}{\rm cn}%
^2(x/k^{\prime }+\alpha \mid k^2).  \label{heqnb}
\end{eqnarray}
We can integrate this equation once by noting that the elliptic integral $E$
satisfies,\cite{byrd}, 
\begin{eqnarray}
\frac 1{k^{\prime }}\frac{{\rm d}}{{\rm d}x}E\left( {\rm am}%
(x/k^{\prime }+\alpha \mid k^2)\right) =1+\left( \frac k{k^{\prime
}}\right) ^2{\rm cn}^2(x/k^{\prime }+\alpha \mid k^2),
\end{eqnarray}
allowing us to write (\ref{heqnb}) as 
\begin{eqnarray}
\frac{{\rm d}h}{{\rm d}x}=\frac 1{k^{\prime }}E\left( {\rm am}%
(x/k^{\prime }+\alpha \mid k^2)\right) -x+C_1.
\end{eqnarray}
A change of variables to ${\rm \xi (x),}$ and the introduction of a new
function $h(x)$ defined by, 
\begin{eqnarray}
h(x)=h_1(x)+\int {\rm d}x(-x+C_1),\;\;\;\;\;\;\xi =am(x/k^{\prime }+\alpha
\mid k^2)
\end{eqnarray}
means that we have 
\begin{eqnarray}
\frac{{\rm d}h_1}{{\rm d}\xi } &=&E(\xi \mid k^2)/\sqrt{1-k^2\sin ^2(\xi
)}  \label{h1eqn} \\
\Rightarrow h_1(\xi ) &=&F^2(\xi \mid k^2)E(k)/(2K(k))+\ln \left( \Theta
[F(\xi \mid k^2)]/\Theta [0]\right) .
\end{eqnarray}
The integration has been done using equation (630.02) of \cite{byrd}. This
further simplifies when we realise \mbox{$F(\xi\mid k^2)=x/k'+\alpha$}. The
conservation of the energy-momentum tensor, \mbox{$T^{\mu\nu}_{\;\;\;\;
;\nu}=0$}, gives the energy density $\epsilon $ as 
\begin{eqnarray}
\epsilon =\left( \beta /(abc)\right) ^2=(\beta /c)^2,  \label{energy}
\end{eqnarray}
where $\beta $ is a constant. The fourth Einstein equation (\ref{int})
reduces to 
\begin{eqnarray}
\ddot h=-\left( (\dot y)^2/4+4\beta ^2\right) .  \label{4einer}
\end{eqnarray}
But we already know $y(t)$ is from (\ref{ysoln}), and the left-hand side
comes from (\ref{heqnb}), so we have the two consistency relations 
\begin{eqnarray}
k^{\prime }{}^2+k^2=1,\;\;\;\;\;k^2/k^{\prime }{}^2=-\beta ^2.
\label{k_beta}
\end{eqnarray}
In what follows, we shall use the value \mbox{$0<\beta<1$}, with %
\mbox{$k=i\beta/\sqrt{1-\beta^2}$} and \mbox{$k'=1/\sqrt{1-\beta^2}$}. Now
that we have a relation between $\beta $ and $k$ we can rewrite (\ref{ysoln}%
) using some elliptic function identities. We also make use of the fact that 
$\alpha $ is an arbitrary integration constant which just shifts the origin
of the solution. It proves convenient to choose \mbox{$\alpha=\sqrt{1-%
\beta^2}K(\beta)$}, turning (\ref{ysoln}) into (see appendix \ref{appAa}) 
\begin{eqnarray}
\cosh (y(t))=1-2\beta ^2{\rm {sn}^2(2t\mid \beta ^2).}  \label{yeqnb}
\end{eqnarray}
This is where we make a connection with the instanton solution. It is seen
that if we take $t$ to be an imaginary parameter, \mbox{$t=i\tau$}, then the
right-hand side of (\ref{yeqnb}) becomes greater than or equal to one,
meaning that $y(\tau )$ is real. 
\begin{eqnarray}
\cosh \left( y(\tau )\right) =1+2\beta ^2{\rm {sn}^2(2\tau \mid 1-\beta ^2)/{%
cn}^2(2\tau \mid 1-\beta ^2).}  \label{yeqnc}
\end{eqnarray}
To proceed then, we are required to do an analytic continuation on (\ref
{h1eqn}) from $t$ to $\tau $ (see appendix \ref{appAb}). 
\begin{eqnarray}
h(\tau ) &=&2\tau ^2(1-E(\beta )/K(\beta ))-\pi \tau ^2/\left( K(\beta )K(%
\sqrt{1-\beta ^2})\right)   \label{Iint0} \\
&&\ \ ~+\ln [\Theta(2\tau\mid 1-\beta^2)]+C_1\tau+C_2  \nonumber
\end{eqnarray}
This completes the solution for the Bianchi VII$_0$ instanton with a stiff
fluid. The solution is defined in the range \mbox{$K(\sqrt{1-\beta^2})<2%
\tau<K(\sqrt{1-\beta^2})$}, with $c(\tau )$ going to zero at these limits.
At these points then we see that the volume of the homogeneous 3-surfaces
goes to zero, defining the end-points of the instanton. It can also be seen
that the curvature of the instanton diverges at these points.

%%%%%%%%%%%%%%%%%%%%%%%%%%%%%%%%%%%%%%%%%%%%%%%%%%%%%%%%%%%%%%%%%%%%%%

\section{Qualitative nature of the solution}

\begin{figure}
\centerline{\setlength\epsfxsize{90mm}\epsfbox{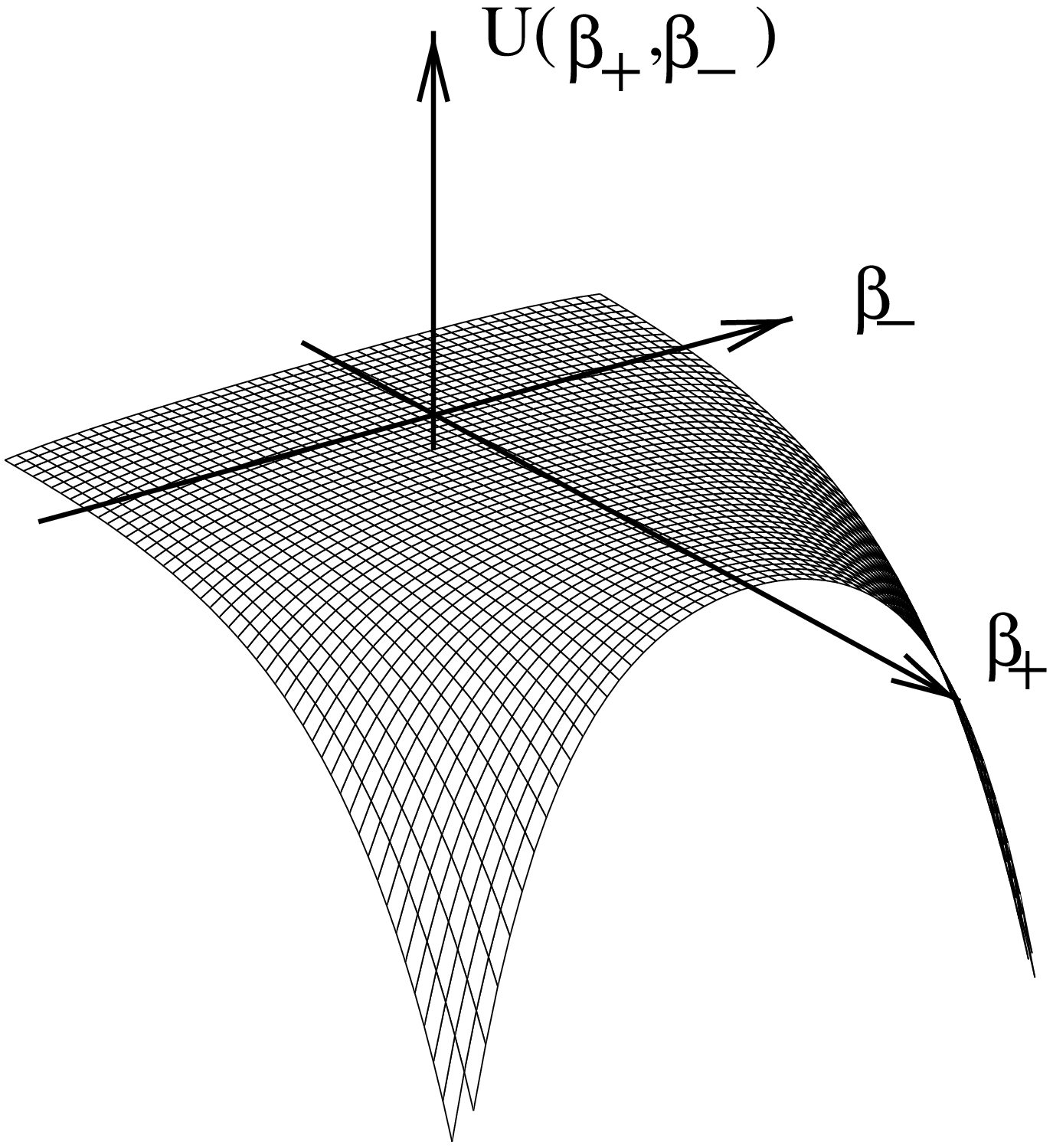}}
 \caption{${\cal U}(\beta_+,\beta_-)$}
 \label{pot}
\end{figure}

\begin{figure}
\centerline{\setlength\epsfxsize{90mm}\epsfbox{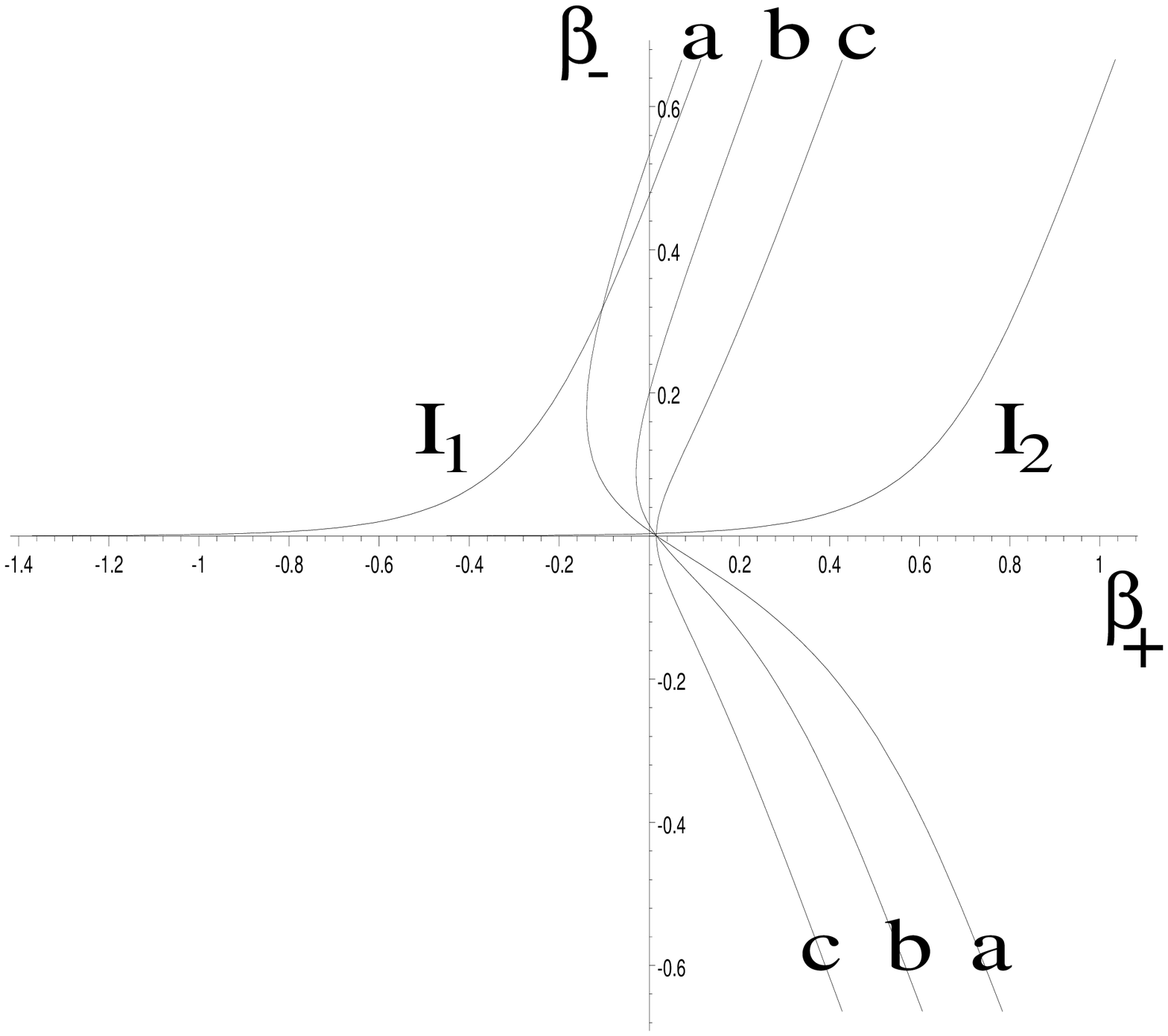}}
 \caption{Various Bianchi VII$_0$ instanton solutions in the
\mbox{$\beta_+-\beta_-$} plane. The I$_{1,\;2}$ are
examples of self dual solutions, with curves a, b, c being
three stiff fluid instantons.}
 \label{bp_bm}
\end{figure}

Let us now examine the system of equations (\ref{aeqn}-\ref{int}) in order
to obtain an effective action. Using imaginary time, $\tau $, we see that
these equations can be derived by extremising the action, 
\begin{eqnarray}
S_{{\rm {eff}}}=\int {\rm d}\tau \left( \alpha ^{\prime }\beta ^{\prime
}+\beta ^{\prime }\gamma ^{\prime }+\alpha ^{\prime }\gamma ^{\prime }-\frac
14(a^2-b^2)^2\right) .  \label{seff}
\end{eqnarray}
Again, we are using \mbox{$\alpha=\ln(a)$}, \mbox{$\beta=\ln(b)$}, %
\mbox{$\gamma=\ln(c)$} and $^{\prime }$ for differentiation with respect to $%
\tau $. The form of this action coincides with the Einstein-Hilbert action
for the Bianchi VII$_0$ metric, found by calculating the curvature of the
connection forms (\ref{connections}) so that this effective action is
proportional to the gravitational action, as expected, 
\begin{eqnarray}
I_{{\rm grav}}}=-\frac 1{16\pi }\int_{{\cal M}}\sqrt{g}R{\rm {d}^4x+%
\textnormal{boundary terms}.
\end{eqnarray}
The boundary terms are designed to cancel the second derivatives coming from
the Ricci scalar term, \cite{gibbons77}, \cite{bmad}, in order that the
equations of motion derive from the variational principle even on a
boundary. The next step is to put the effective action (\ref{seff}) into a
familiar form, so we can visualise out solution. This is achieved by using
the following Misner variables \cite{gibbons79}, 
\begin{eqnarray}
a(\tau ) &=&\exp (\Omega (\tau )+\beta _{+}(\tau )+\sqrt{3}\beta _{-}(\tau ))
\nonumber  \label{newparams} \\
b(\tau ) &=&\exp (\Omega (\tau )+\beta _{+}(\tau )-\sqrt{3}\beta _{-}(\tau ))
\\
c(\tau ) &=&\exp (\Omega (\tau )-2\beta _{+}(\tau ))  \nonumber
\end{eqnarray}
We then find that the effective action, in terms of the new functions, takes
the familiar form for a particle moving in a potential well, (\ref{newSeff})
\begin{eqnarray}
S_{{\rm {eff}}}=-3\int {\rm {d}\tau \left( \beta _{+}^{\prime }{}^2+\beta
_{-}^{\prime }{}^2-\Omega ^{\prime }{}^2+\frac 13\exp (4\Omega )\exp (4\beta
_{+})\sinh ^2(2\sqrt{3}\beta _{-})\right) }  \label{newSeff}
\end{eqnarray}
The $\beta _{\pm }$ functions evolve to create the time-dependent
potential
\newline \mbox{${\cal V}(\Omega,\beta_+,\beta_-)=\exp(4\Omega){\cal U}(\beta_+,\beta_-)$}. 
Fig. \ref{pot} illustrates the $\Omega $-independent
part of the potential, with a ridge along the $\beta _{+}$ axis. The
behaviour of the two solutions outlined above, the self-dual and stiff fluid
instantons, are given in Fig. \ref{bp_bm}. The figure shows the self-dual
solution (paths I$_1$ and I$_2$) evolves along the ridge and then falls off
at some point determined by the arbitrary parameter $\lambda $ of the
solution, (\ref{sd_metric}). The solution for the stiff fluid contains three
parameters, $k$, $C_1$ and $C_2$.The general behaviour is shown in Fig. \ref
{bp_bm}, which shows the solution evolving toward and then rolling over the
ridge.

%%%%%%%%%%%%%%%%%%%%%%%%%%%%%%%%%%%%%%%%%%%%%%%%%%%%%%%%%%%%%%%%%%%%%%

\section{Bianchi VI$_0$ solution}

\begin{figure}
\centerline{\setlength\epsfxsize{90mm}\epsfbox{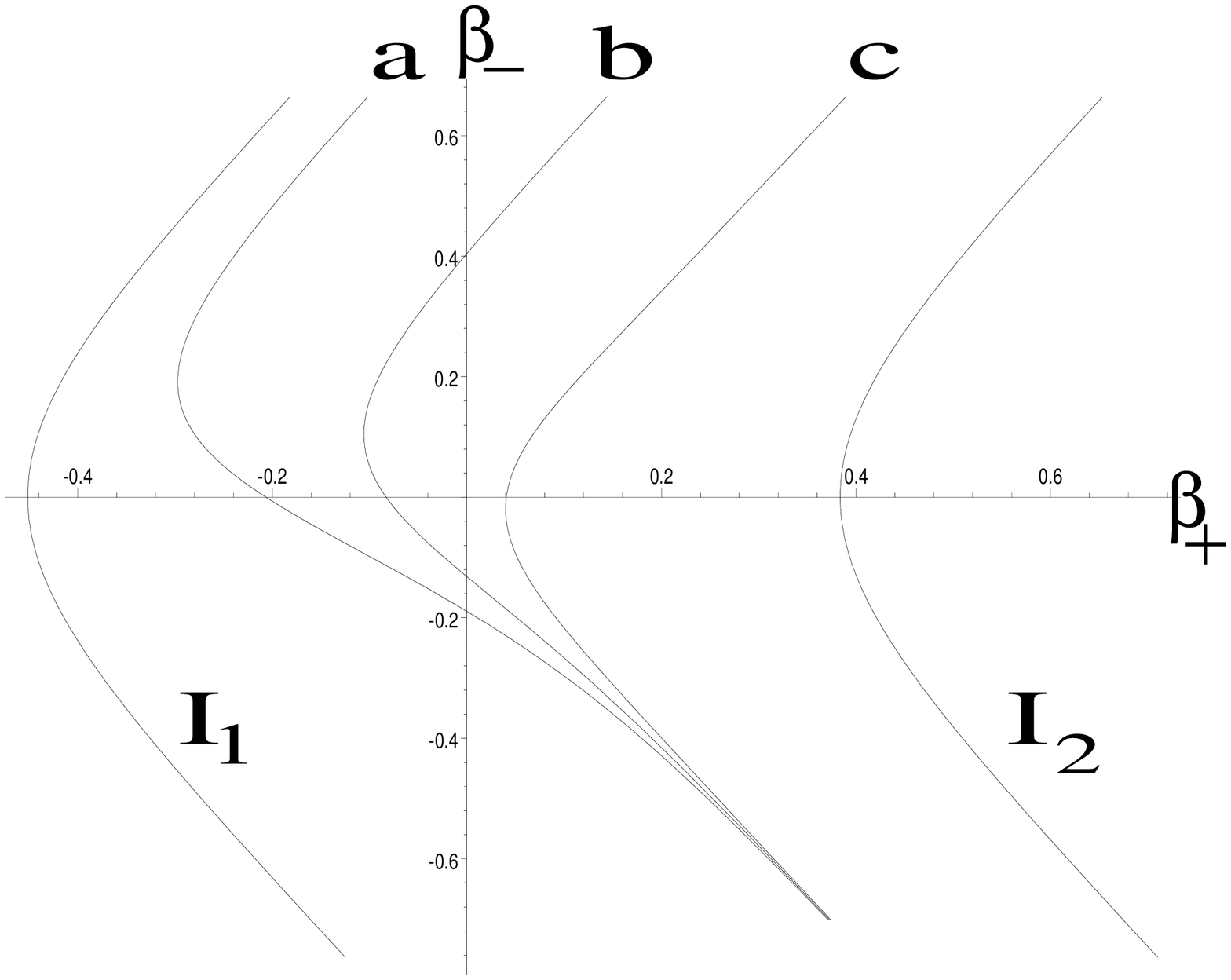}}
 \caption{Various Bianchi VI$_0$ instanton solutions in the
\mbox{$\beta_+-\beta_-$} plane. The I$_{1,\;2}$ are
examples of self dual solutions, with curves a, b, c being
three stiff fluid instantons.}
 \label{bp_bm6}
\end{figure}

The Bianchi VI$_0$ structure constants differ from Bianchi VII$_0$ (\ref
{b7struct}) by a single minus sign: 
\begin{eqnarray}
C_{\;32}^1=-C_{\;23}^1=-C_{\;13}^2=C_{\;31}^2=1.  \label{b6struct}
\end{eqnarray}
This changes the symmetry group from the two dimensional Euclidean group,
E(2), to E(1,1), and also changes one of the equations of motion, (\ref{ceqn}%
). Using the same metric as (\ref{orth}), but with the left-invariant one
forms of E(1,1), we find the new equation, 
\begin{eqnarray}
\left( \ln (c^2)\right) ^{..}-(a^2-b^2)^2 &=&0.  \label{c6eqn}
\end{eqnarray}
Again we make the ansatz \mbox{$a(t)=1/b(t)$} giving 
\begin{eqnarray}
\frac{{\rm d}^2h}{{\rm d}x^2}=1+\sinh ^2(y/2),  \label{heqn6}
\end{eqnarray}
where we have again taken \mbox{$x=2t$}, \mbox{$a^2=\exp(y/2)$} and %
\mbox{$c^2=\exp(h)$}. Comparing equations (\ref{ay}) and (\ref{heqn6}), we
see that the solution for $h(x)$ in the Bianchi VI$_0$ case is the same as
for the Bianchi VII$_0$ case, except that there is an addition of $\frac
12x^2$ . However, the Bianchi VI$_0$ solution differs in other respects. By
looking at the fourth Einstein equation, (\ref{ein4}), we find a different
relation between $k$, $k^{\prime }$ and the parameter $\beta $ of (\ref
{energy}). Using (\ref{4einer}) and the new $h(t)$ we may derive, 
\begin{eqnarray}
k^{\prime }{}^2+k^2=1,\;\;\;\;\;k^2/k^{\prime }{}^2=-(1+\beta ^2).
\label{k_beta6}
\end{eqnarray}
Using methods similar to the previous analysis one may then derive, 
\begin{eqnarray}
\cosh (y(\tau )) &=&1+2(1+\beta ^2){\rm cs}^2\left( 2\sqrt{1+\beta ^2}\tau
\mid \frac{\beta ^2}{1+\beta ^2}\right)  \\
h(\tau ) &=&-2\left( 1-E(\beta /\sqrt{1+\beta ^2})/K(\beta /\sqrt{1+\beta ^2}%
)\right) \tau ^2\\
\nonumber &~&+\ln \;\vartheta _1\left( \frac{\sqrt{1+\beta ^2}\pi }{%
K(\beta /\sqrt{1+\beta ^2})}\tau \mid \frac{\beta ^2}{1+\beta ^2}\right)
+C_1\tau +C_2
\end{eqnarray}
This instanton is defined in the range \mbox{$0<\tau<E(\beta/\sqrt{1+%
\beta^2})/\sqrt{1+\beta^2}$}. Following the procedure outlined for the
Bianchi VII$_0$ case, we may also derive the self-dual solution, 
\begin{eqnarray}
{\rm {d}s^2=\frac{\lambda ^2}2\sin (2\tau )\left( {d}\tau ^2+(\sigma
^3)^2\right) +\cot (\tau )(\sigma ^1)^2+\tan (\tau )(\sigma ^2)^2,}
\label{sd_metric6}
\end{eqnarray}
The Hamiltonian method used to describe the qualitative behaviour may also
be adapted. On doing so we find that the solution is described by the
effective action as (\ref{newSeff}), only now the $\sinh ^2(2\sqrt{3}\beta
_{-})$ gets replaced by $\cosh ^2(2\sqrt{3}\beta _{-})$, using the same
allocations as (\ref{newparams}). The effective potential for the $\beta
_{\pm }$ is thus qualitatively the same as Fig. \ref{pot}. The paths that
this solution describes in the $\beta _{\pm }$ plane are shown in Fig. \ref
{bp_bm6}, where the examples of self-dual solutions are labelled $I_1$ and $%
I_2$, with $a$, $b$ and $c$ being stiff fluid examples. 
%%%%%%%%%%%%%%%%%%%%%%%%%%%%%%%%%%%%%%%%%%%%%%%%%%%%%%%%%%%%%%%%%%%%%%

\section{Bianchi VII$_0$ solution with electromagnetic field}

In this section we give an analytic solution of Bianchi type VII$_0$ in the
presence of an electromagnetic two-form satisfying Maxwell's relations.
Using the orthonormal basis for the metric we write Maxwell's relations as, 
\begin{eqnarray}
{\rm {dF}=0,\;\;\;{d}^{*}{F}=0}  \label{maxwell}
\end{eqnarray}
which are valid here because there are no charged sources. We consider here
a two-form field strength of the form 
\begin{eqnarray}
{\rm {F}=E\omega ^0\wedge \omega ^3+B\omega ^1\wedge \omega ^2,}
\end{eqnarray}
representing an electric and a magnetic field in the `3' direction. One
quickly sees then that (\ref{maxwell}) leads to both $E$ and $B$ being
constant for the Bianchi VII$_0$ metric. When it comes to the field
equations for the metric coefficients we have only a minor change. The
right-hand side of (\ref{aeqn}-\ref{ceqn}) no longer vanishes but becomes $%
\eta Uc^2$, where $\eta $ is $+1$ for (\ref{aeqn}), (\ref{beqn}) and $-1$
for (\ref{ceqn}). The constant $U$ that appears is given by %
\mbox{$U=(E)^2+(B)^2$} (see \cite{spokoiny}). There is a clear choice of
ansatz for these equations, namely the biaxial ansatz with \mbox{$a=b$}.
Taking this ansatz leads to 
\begin{eqnarray}
(\ln a^2)^{\prime \prime } &=&-Uc^2  \label{c1} \\
(\ln c^2)^{\prime \prime } &=&Uc^2,
\end{eqnarray}
and with the definition \mbox{$c^2=\exp(\theta)$} the equation for $c$ 
gives, 
\begin{eqnarray}
\theta ^{\prime \prime }=U\exp (\theta ),  \label{th1}
\end{eqnarray}
taking note that we are now using Euclidean time, \mbox{$\tau=it$}. We can
trivially perform the first integral of (\ref{th1}) by multiplying both
sides by $\theta ^{\prime }$ giving, 
\begin{eqnarray}
\int \frac{{\rm {d}\theta }}{\sqrt{D^2+2U\exp (\theta )}}=\int {\rm {d}\tau .%
}
\end{eqnarray}
Integrating this and using (\ref{c1}) we find 
\begin{eqnarray}
c^2(\tau ) &=&\frac{(x^2-D^2)}{2U} \\
a^2(\tau ) &=&\exp (C_1\tau +C_2)\frac{2U}{x^2-D^2} \\
x &=&D\frac{1+\exp (\pm D(\tau +\tau _0))}{1-\exp (\pm D(\tau +\tau _0))},
\end{eqnarray}
with $C_1$, $C_2$, $D$ and $\tau _0$ all constants of integration. The 4th
Einstein field equation (eq.(15) in \cite{spokoiny}) gives the constraint $%
C_1=D$. The qualitative behaviour of this electromagnetic Bianchi VII$_0$
model follows from the previous discussion. The extra terms $\eta Uc^2$ in
the equations of motion correspond to the addition of an extra term $\frac
12Uc^2$ in the effective action (\ref{seff}), which has the effect of
tilting the potential in (\ref{newSeff}) in the $\beta _{+}$ direction. The
trajectory in \mbox{$(\beta_+,\; \beta_-)$} space is clear in this biaxial
example, \mbox{$\beta_- =0$} so the solution rolls along the axis.

%%%%%%%%%%%%%%%%%%%%%%%%%%%%%%%%%%%%%%%%%%%%%%%%%%%%%%%%%%%%%

\section{Conclusions}

Exact solutions of the Einstein equations are only possible in the 
presence of symmetries or structural specialisations. When the metric 
signature is Riemannian it is possible to find vacuum solutions when 
self duality is combined with three-dimensional homogeneity. We have 
used the structure of these solutions to find a number of non-vacuum 
solutions with stiff matter and electromagnetic field sources when 
self duality is necessarily relaxed.  These solutions are of Bianchi 
types VI and VII. 

%%%%%%%%%%%%%%%%%%%%%%%%%%%%%%%%%%%%%%%%%%%%%%%%%%%%%%%%%%%%%%%%%%%%%%%

\section*{Acknowledgements}

We would like to acknowledge helpful discussions with Gary Gibbons and
Stephen Siklos. JDB was supported supported by a PPARC Senior
Fellowship, and PMS was partially funded by PPARC.

%%%%%%%%%%%%%%%%%%%%%%%%%%%%%%%%%%%%%%%%%%%%%%%%%%%%%%%%%%%%%%%%%%%%%%%%%%

%%%%%%%%%%%%%%%%%%%%%%%%%%%%%%%%%%%%%%%%%%%%%%%%%%%%%%%%%%%%%%%%%%%%%
\appendix

\section{Appendix}

\label{appA}

\label{appAa} In order to transform (\ref{ysoln}) to (\ref{yeqnc}), 
\begin{eqnarray}
\cosh (y)=1-2\beta ^2{\rm {cn}^2\left( 2t/k^{\prime }+{\alpha }\mid -\frac{%
\beta ^2}{1-\beta ^2}\right) }
\end{eqnarray}
we use the transformation of elliptic functions for negative parameter, as
found in \cite{abramowitz}. 
\begin{eqnarray}
\cosh (y)=1-2\beta ^2{\rm {cd}^2\left( (2t/k^{\prime }+{\alpha })/\sqrt{%
1-\beta ^2}\mid \beta ^2\right) .}
\end{eqnarray}
We use the fact that $\alpha $ is arbitrary and define, \mbox{$\alpha/%
\sqrt{1-\beta^2}=K(\beta)$}, and find 
\begin{eqnarray}
\cosh (y)=1-2\beta ^2{\rm {sn}^2(2t\mid \beta ^2).}
\end{eqnarray}
The final step is to perform the continuation \mbox{$t=i\tau$} which gives
us (\ref{yeqnc}).

\label{appAb} Here we continue equation (\ref{h1eqn}) to imaginary time %
\mbox{$t=i\tau$}, 
\begin{eqnarray}
h(t)=(2t/k^{\prime }+\alpha )^2E(k)/(2K(k))+\ln \left( \Theta [2t/k^{\prime
}+\alpha \mid k^2]/\Theta [0]\right) -2t^2+C_1t+C_2.
\end{eqnarray}
We use the definition given by Cayley of the theta function, \cite{cayley}, 
\begin{eqnarray}
\frac{\Theta (u\mid k^2)}{\Theta (0\mid k^2)} &=&\exp \left[\frac
12u^2(1-E(k)/K(k))-k^2\int_0^u{\rm d}z\int_0^z{\rm d}y{\rm sn}^2(y\mid k^2)%
\right]   \label{heqn2} \\
\Rightarrow h(t) &=&\frac 12(2t/k^{\prime }+\alpha
)^2-2t^2+C_1t+C_2\\
\nonumber &~&-k^2\int_0^{2t/k^{\prime }+\alpha} {\rm d}z\int_0^z{\rm d}y\;{\rm sn}^2(y\mid k^2)
\end{eqnarray}
Define the double integral in (\ref{heqn2}) to be, 
\begin{eqnarray}
I_1=\int_0^{2t/k^{\prime }+\alpha }{\rm d}z\int_0^z{\rm d}y\;{\rm sn}^2(y\mid k^2).
\end{eqnarray}
Change $t$ to $i\tau $ and make two changes of variables, %
\mbox{$z=2iv/k'+\alpha$} and \mbox{$y=2i\xi/k'+\alpha$} to obtain 
\begin{eqnarray}
I_1(\tau )=A\tau +B-4/k^{\prime }{}^2\int_0^\tau {\rm {d}v\int_0^v{d}\xi
\frac 1{{cn}^2(2\xi \mid 1-\beta ^2)}.}
\end{eqnarray}
In deriving this we have used the relation between $k$, $k^{\prime }$ and $%
\beta $ (\ref{k_beta}); note that $k$ is imaginary. The double integral can
be rewritten in terms of theta functions by making the following change of
variables, \mbox{$\xi=iu/2$} and \mbox{$v=iy/2$}. This leaves us with a
double integral which may be related to a theta function with imaginary
argument. This can be expressed in terms of a theta function with real
argument using a result in \cite{cayley}, and hence we obtain (\ref{Iint0}).

%%%%%%%%%%%%%%%%%%%%%%%%%%%%%%%%%%%%%%%%%%%%%%%%%%%%%%%%%%%%%%%%%%%%%%%%%%

%\begin{figure}[!htb]
%\centering
%\mbox{\epsfig{file=U_omega_fig.eps,height=3.0in,width=3.0in}}
%          \caption{${\cal U}(\beta_+,\beta_-)$}
%          {
%          
%          }
%\label{pot}
%\end{figure}

%\begin{figure}[!htb]
%\centering
%\mbox{\epsfig{file=soln_fig.eps,height=3.0in,width=4.5in}}
%          \caption{Various Bianchi VII$_0$ instanton solutions in the
%\mbox{$\beta_+-\beta_-$} plane. The I$_{1,\;2}$ are
%examples of self dual solutions, with curves a, b, c being
%three stiff fluid instantons.}
%          {
          
%          }
%\label{bp_bm}
%\end{figure}

%\begin{figure}[!htb]
%\centering
%\mbox{\epsfig{file=soln6_fig.eps,height=3.0in,width=4.5in}}
%          \caption{Various Bianchi VI$_0$ instanton solutions in the
%\mbox{$\beta_+-\beta_-$} plane. The I$_{1,\;2}$ are
%examples of self dual solutions, with curves a, b, c being
%three stiff fluid instantons.}
%          {
%          
%          }
%\label{bp_bm6}
%\end{figure}
\end{document}